\documentclass[sigconf]{acmart}

\usepackage{booktabs} 
\usepackage{longtable}

\setcopyright{none}

\begin{document}
\title{Key Challenges with Agile Culture - A Survey among Practitioners}
\titlenote{Produces the permission block, and
  copyright information}

\renewcommand{\shorttitle}{Key Challenges with Agile Culture - A Survey among Practitioners}





\author{Thorben Kuchel}
\affiliation{%
  \institution{Bagilstein GmbH}
  \city{Mainz} 
  \country{Germany} 
}
\email{thorben.kuchel@bagilstein.de}

\author{Michael Neumann}
\affiliation{%
  \institution{University of Applied Sciences and Arts Hannover}
  \department{Dpt. of Business Information Systems}
  \city{Hannover} 
  \country{Germany} 
}
\email{michael.neumann@hs-hannover.de}

\author{Philipp Diebold}
\authornote{Also with IU Unternational University, Erfurt, Germany, philipp.diebold@iu.org}
\affiliation{%
  \institution{Bagilstein GmbH}
  \city{Mainz} 
  \country{Germany} 
}
\email{philipp.diebold@bagilstein.de}

\author{Eva-Maria Schön}
\affiliation{%
  \institution{University of Applied Sciences and Arts Emden/Leer}
  \department{Dpt. of Economics}
  \city{Emden}
  \country{Germany}
}
\email{eva-maria.schoen@hs-emden-leer.de}

\renewcommand{\shortauthors}{Kuchel et al.}

\begin{abstract}
\textit{Context:} Within agile transformations, there are a lot of different challenges coming up. One very important but less considered and treated in research are cultural challenges. Although research shows that cultural clashes and general organizational resistance to change are part of the most significant agile adoption barriers.  
\textit{Objective:} Thus, our objective is to tackle this field and come up with important contributions for further research. To this end, we want to identify challenges that arise from the interplay
between agility and organizational culture. 
\textit{Method:} This is done based on an iterative research approach. On the one hand, we gathered qualitative data among our network of agile practitioners. Then, we derived in sum 15 challenges with agile culture. On the other hand, we gathered quantitative data by means of a questionnaire study with 92 participants.     
\textit{Results:} We identified 7 key challenges out of the 15 challenges with agile culture. The results that are presented in a conceptual model show a focus on human aspects that we need to deal with more in future.
\textit{Conclusion:} Based on our results, we started deriving future work aspects to do more detailed research on the topic of cultural challenges while transitioning or using agile methods in software development and beyond.
\end{abstract}

%
%

\ccsdesc[500]{Software and its engineering~Agile software development}

\keywords{Agile software development, agile culture, organizational culture, challenges}

\maketitle

\section{Introduction}

Agile methods are established approaches in software development for more than 20 years \cite{Abrahamsson.2002}. Today, many companies worldwide use agile methods in order to be able to react to rapidly changing conditions. The iterative approach and a high degree of communication and interaction make it possible to react adequately to changing requirements and individual customer and user needs. This enables companies to be competitive even in highly volatile market environments and to be able to face the VUCA world \cite{Bennett.2014}. Due to the strong focus on social aspects such as communication, interaction and integration of stakeholders, values and principles are of great importance when introducing and using agile methods in practice~\cite{Chow.2008}. In addition to the agile manifesto~\cite{Beck.2001}, in which four pairs of values and 12 principles are presented, specific values are described in the guidelines for agile methods such as Scrum~\cite{Schwaber.2020}. 

However, we know that many companies are challenged by their agile transition~\cite{VersionOne.2021}. The successful transition and use of agile methods in practice depends on various factors \cite{Diebold.2022}. In addition to aspects such as the successful involvement of stakeholders \cite{Hoda.2011}, the correct transition and application of agile practices among the different organizational levels is important~\cite{Neumann.2022}. Furthermore, the success of agile methods depends on cultural aspects of the respective organization~\cite{Diebold.2015}. The State of Agile Report \cite{VersionOne.2021} presents the contradictions between the organizational culture and agile values as one of the core challenges in the agile transition. For instance, the prevailing error culture in the company is important to actively promote the self-organization of agile teams. In order to achieve the necessary degree of transparency in one’s work and critical questioning of the agile approach in use, trust between the team members is necessary\cite{Schoen.2015}. 

The key challenges that arise from cultural aspects when introducing and applying agile methods in practice are diverse~\cite{Gelmis.2022}. So far, this topic has been underrepresented in the recent literature (see Section~\ref{Sec2:RelWork}). Based on the results of our iterative research approach, we will answer our research question: \textit{What are the key challenges with agile culture?}

\begin{table*}
 \caption{Overview of the related work}
  \label{tab1:OverviewofRelWork}
   \begin{tabular}{ccp{0.5\linewidth}}
\hline
Publication & Cultural level(s) & Findings \\
\hline
Gelmis et al. (2022)~\cite{Gelmis.2022} & National culture & Hierarchical cultural structures, which are described for Turkish culture, are challenging the use and transition of agile methods. The authors recommend transforming to a more flat organizational structure, which supports the use of agile methods and should optimize the performance of agile teams.\\
\hline
Gupta et al. (2019)~\cite{Gupta.2019} & Organizational culture & The results of the study show that a hierarchical culture hinders the use of social and technical agile practices. In contrast, the authors were able to show that a group culture has a positive influence on the application of social agile practices. Development cultures, in turn, support both social and technical agile practices. Based on the results of their study, the authors point out the many facets of culture and the respective influences on a successful agile transition of IT departments in organizations. They therefore recommend considering the underlying cultures in the organization before starting an agile transition. For this you describe the six-step process described by Cameron \& Quinn~\cite{Cameron.2006}.\\
\hline
Ivari \& Ivari (2011)~\cite{Iivari.2011} & Organizational culture & Based on an empirical study, the author points out that, for example, the hierarchical culture is incompatible with the use of agile methods. For example, it is shown that the transition and use of agile methods in hierarchical organizational cultures leads to a steadily increasing formalization of the agile method and that it can lose key elements of agile work over time.\\
\hline
Siakas \& Siakas (2007)~\cite{Siakas.2007} & National and organizational culture & In their work, the authors describe an agile culture based on well-known success factors such as user satisfaction and stakeholder involvement. They present the connections between organizational and cultural aspects and elements of agile working such as practices and roles. The authors also point out the importance of employee motivation and dynamics and draw parallels to established frameworks such as ETHICS \cite{Mumford.2000} or TQM \cite{Deming.1986}.\\
\hline
Strode et al. (2009)~\cite{Strode.2009} & Organizational culture & Based on their empirical study, the authors present specific correlations between organizational culture characteristics and the successful application of agile methods. For example, they point out the importance of positive evaluation of feedback and learning. They also name the trusting social interaction, a collaborative and competent cooperation and the result orientation of the organization.\\
\hline
Tolfo et al. (2011)~\cite{Tolfo.2011} & Organizational culture & The authors address the supporting and hindering influences of organizational culture on the transition of agile methods. The authors point out that specific influences manifest themselves on the specific sub-levels (e.g. on team level) of the organizational culture. A superficial examination of these influences can be a hindrance and lead to incorrect measures. Rather, it is important to consider the different levels of organizational culture in order to be able to develop an agile culture in the long term.\\
\hline
\end{tabular}
\end{table*}

The paper at hand is structured as follows: We outline the related work in Section \ref{Sec2:RelWork}. In Section \ref{Sec3:ResearchDesign}, we provide an overview of our iterative research approach. The results are presented and discussed in Section \ref{Sec4:Results}, including the limitations of our study. Finally, we give a conclusion and describe the future work in Section \ref{Sec5:ConclusionAndFutureWork}. 

\section{Related Work}
\label{Sec2:RelWork}

We searched for both, primary and secondary studies in order to provide an overview of the literature related to key challenges of cultural aspects while transitioning or using agile methods or practices. The search was performed using Google Scholar. Table \ref{tab1:OverviewofRelWork} gives a brief overview of the identified related work.

\begin{figure*}
\caption{Overview iterative research approach}
\label{Figure1}
\includegraphics[scale=0.50]{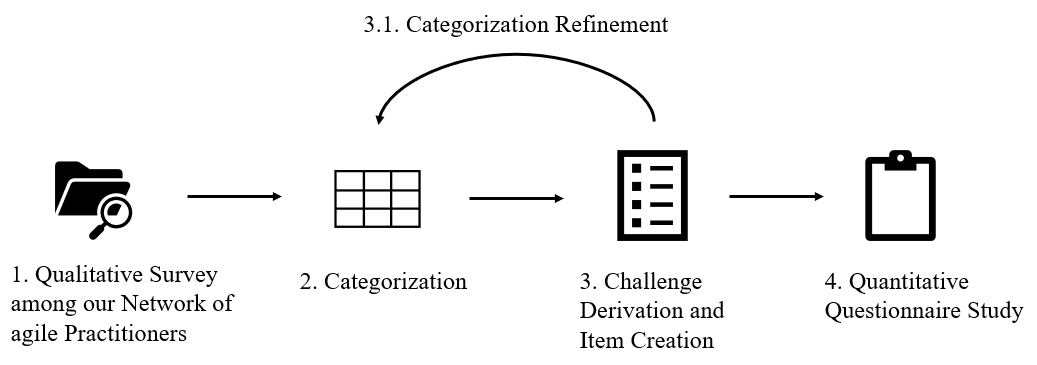}
\end{figure*}

Various authors point to the multiple facets of culture \cite{Gupta.2019,Siakas.2007}. The overview of the related work also shows that different cultural levels (country, regions, companies, departments, teams and individuals) may have different influences on the transition and use of agile methods and practices. In addition, agile methods can be very diverse in practice due to the tailoring of the respective method~\cite{Diebold.2014,Neumann.2022}. Most of the articles we identified address the influence of organizational culture on the transition or use of agile methods and practices. Interestingly, different levels of detail are considered here. While some use organizational culture and its characteristics on an abstract level \cite{Iivari.2011,Siakas.2007}, others try to make a concrete reference to teams and individuals \cite{Tolfo.2011}.

Although we tried to find related work dealing with key challenges of cultural aspects while using agile methods and practices, we could not identify literature dealing closely with the aim of our study. So, to the best of our knowledge, this is the first publication that identifies key challenges with agile culture. 

\begin{table}
 \caption{Example of derivation of a challenge (see Step 3)}
  \label{tab2:Example of derivation}
  \begin{tabular}{cp{2.2cm}p{2.2cm}p{2.2cm}}
\hline
ID & Problem statement & Item (1. Iteration) & Item (Final) \\
\hline
C5 & The existing structures of an organization prevent an agile way of working. & How important is it to create structures that support an agile way of working? & How important is it to create organizational structures that support an agile way of working?\\
\hline
\end{tabular}
\end{table}

\section{Research Design}
\label{Sec3:ResearchDesign}

This paper aims to identify challenges that arise from the interplay between agility and organizational culture. To this end, we want to address the following research question: \textit{What are the key challenges with agile culture?}
To achieve our objective, we applied an iterative research approach (see Figure~\ref{Figure1}). At each step, we gathered feedback from practitioners and discussed it extensively among the group of authors.

\paragraph{Step 1}: In the beginning we started with a qualitative survey among our network of agile practitioners in order to gather input for our quantitative questionnaire. Therefore, the third author of the paper asked the question at the professional network LinkedIn: “What do you see as tangible problems with agile and culture?” In sum, we received 34 qualitative answers from 28 participants.   

\paragraph{Step 2}: These answers were categorized by a coding scheme, which ended up with 15 groups in Excel (agile values, willingness to change, fixed structures and hierarchies, agile leadership, technical vs. cultural agility, respect, transparency of decisions, transparency of processes, trust, perseverance, feedback culture, experiments, failure culture, comfort zone and flight levels). This preliminary result was intensively discussed among the authors and then refined where appropriate.  

\paragraph{Step 3}: In the next step, we derived items for our quantitative questionnaire. First, we formulated problem statements and reformulated them into questions. These items were discussed and refined among the authors in several iterations. An example that illustrates this process is shown in Table \ref{tab2:Example of derivation}.

\paragraph{Step 4}: Then we conducted the quantitative questionnaire study. Accordingly, we set up an online survey using Google forms. We used appropriate guidelines for developing our questionnaire \cite{Graf.2002,Finstad.2010,Runeson.2009} and pre tested it with five participants. In sum, the questionnaire comprises 21 items. Five of those items queried socio-demographic data which helps us to describe our sample, 15 items queried the potential key challenges with agile culture (see C1-C15 in Appendix A.1) and one item posed additional comments. The participants of the survey rated the potential key challenges with agile culture (see C1-C15 in Appendix A.1) in terms of importance using a 7-point Likert scale (totally unimportant, unimportant, rather unimportant, neutral, rather important, important, and totally important). In addition, they had the option to give no statement. 

\subsection{Data Collection and Analysis}
In the following, we want to present more details regarding the data collection and analysis of our quantitative questionnaire (see Step 4). The quantitative questionnaire was online between 11/26/2021 and 12/17/2021. It was spreaded via the professional network LinkedIn and personal contacts of the authors. In sum, we received 93 fully completed datasets. No participant terminated the questionnaire before the end. So, the dropout rate is 0\%. This surprised us, but can be explained by the good design of the questionnaire, oriented towards the target group. However, we removed one set of data from the sample because one participant showed unusual response behavior. The unusual response behavior was identified by analyzing the data set on detail. We found one respondent which selected the same value of the likert scale for all questions except one.  

Furthermore, we used statistical data (mean, standard deviation and confidence interval) for the analysis of the data. We defined challenges as key in those cases where 50\% of the participants weighted them as totally important. Based on this interpretation, we were able to identify a total of seven key challenges (see Table \ref{tab3:Key Challenges with AC}).

\subsection{Description of the Sample}
The participants of the quantitative questionnaire (sample size N=92) had between 0 to 30 years (mean 6.13) of experience with an agile way of working. Of the 92 participants (N=92), 74 work in the private sector, 9 in the public sector, 10 work at universities or research institutes while 18 are self-employed (multiple answers were possible). Moreover, participants work in different industries like IT, consulting, ecommerce, research and development, automobile, logistics, eHealth, insurance, energy, and social. We asked the participants (N=92) for their general opinion on agility and they answered as follows: “overrated, a buzzword to me” (2,2\%), “should be applied where appropriate” (63\%), “urgently needed in today's world” (33,7\%), and “neutral” (1,1\%). Summarizing this information, we can observe that on the one hand, the sample has a good mix in terms of their experience with agile ways of working as well as working in different industries. On the other hand, the sample is not opposed to agile ways of working.

\section{Results and Discussion}
\label{Sec4:Results}
This section presents the results of our survey. First, we will present the identified key challenges with agile culture that practitioners face today. Then, we will discuss the implications of these challenges on practitioners and researchers. Finally, we will outline the limitations of this research.

\subsection{Key Challenges with Agile Culture}
We identified seven key challenges that arise from the interplay between agility and organizational culture (see Table~\ref{tab3:Key Challenges with AC}). The response options of the Likert scale were normalized for evaluation to the value range 1 (totally unimportant) < x < 7 (totally important). The key challenges in Table~\ref{tab3:Key Challenges with AC} were rated as totally important by more than 50\% of the participants in our survey. In addition, the importance of all key challenges is between important and totally important, which is shown by the small confidence interval. The statistical data for all queried challenges is presented in Appendix A.1.

\begin{table}
 \caption{Key Challenges with Agile Culture}
  \label{tab3:Key Challenges with AC}
  \begin{tabular}{p{0.3cm}p{2.5cm}p{0.65cm}p{0.75cm}p{0.9cm}p{0.75cm}p{0.3cm}}
\hline
ID & Item (EN) & Mean & Stand. deviation & Confid. (p=0.05) & totally important & N \\
\hline
C1 & Humans in an organization do not treat each other with respect. & 6.70 & 0.81 & 0.17 & 80.4\% & 92\\
\hline
C2 & Management expects change from employees without embodying agile values themselves. & 6.43 & 1.17 & 0.24 & 69.6\% & 92\\
\hline
C3 & The organizational culture does not create a context for trusting interactions. & 6.60 & 0.63 & 0.13 & 67.4\% & 92\\
\hline
C4 & Humans in an organization are not allowed to make mistakes. & 6.49 & 0.75 & 0.15 & 60.9\% & 92\\
\hline
C5 & The existing structures of an organization prevent an agile way of working. & 6.40 & 0.96 & 0.20 & 59.8\% & 92\\
\hline
C6 & Agile cultural change does not occur at all levels (individual, team, management) of the organization. & 6.23 & 1.12 & 0.23 & 53.3\% & 90\\
\hline
C7 & Feedback is not valued in an organization. & 6.35 & 0.76 & 0.16 & 51.1\% & 92\\
\hline
\end{tabular}
\end{table}

If we take a deeper dive into the topics of each key challenge, we can observe that there are issues regarding internalizing and living agile values (C2), and especially  values like respect (C1) and trust (C3, C4, and C7). In addition, the data highlights issues with existing structures in organizations (C5) that prevent an agile way of working as well as issues with a cultural change at different levels (individual, team, management) of an organization (C6). 

Figure 2 outlines the relationships between the key challenges with agile culture in a conceptual model. Fixed structures, established thought patterns and strict hierarchies with command \& control are a major barrier to agile ways of working. Organizational structures interact with the involvement of all levels (management, team, individual) and organizations should adapt their structures iteratively to optimize those interactions between levels. In doing so, it is important to also consider the interaction with the organizational culture. Moreover, organizational structures interact with the management level. The change from boss to leader culture \cite{Krieg.2022} represents a key factor for more agility by living out values, behaviors, and a culture of open feedback and learning to establish a growth mindset \cite{Dweck.2017} among employers. We understand boss as a special characteristic of management including aspects like a hierarchical understanding of an organization or a top down approach in decision making. From our point of view, it is essential that a modern leader acts as a role model for agile teams in an organization. 

\begin{figure*}
\label{Figure2}
\includegraphics[scale=0.35]{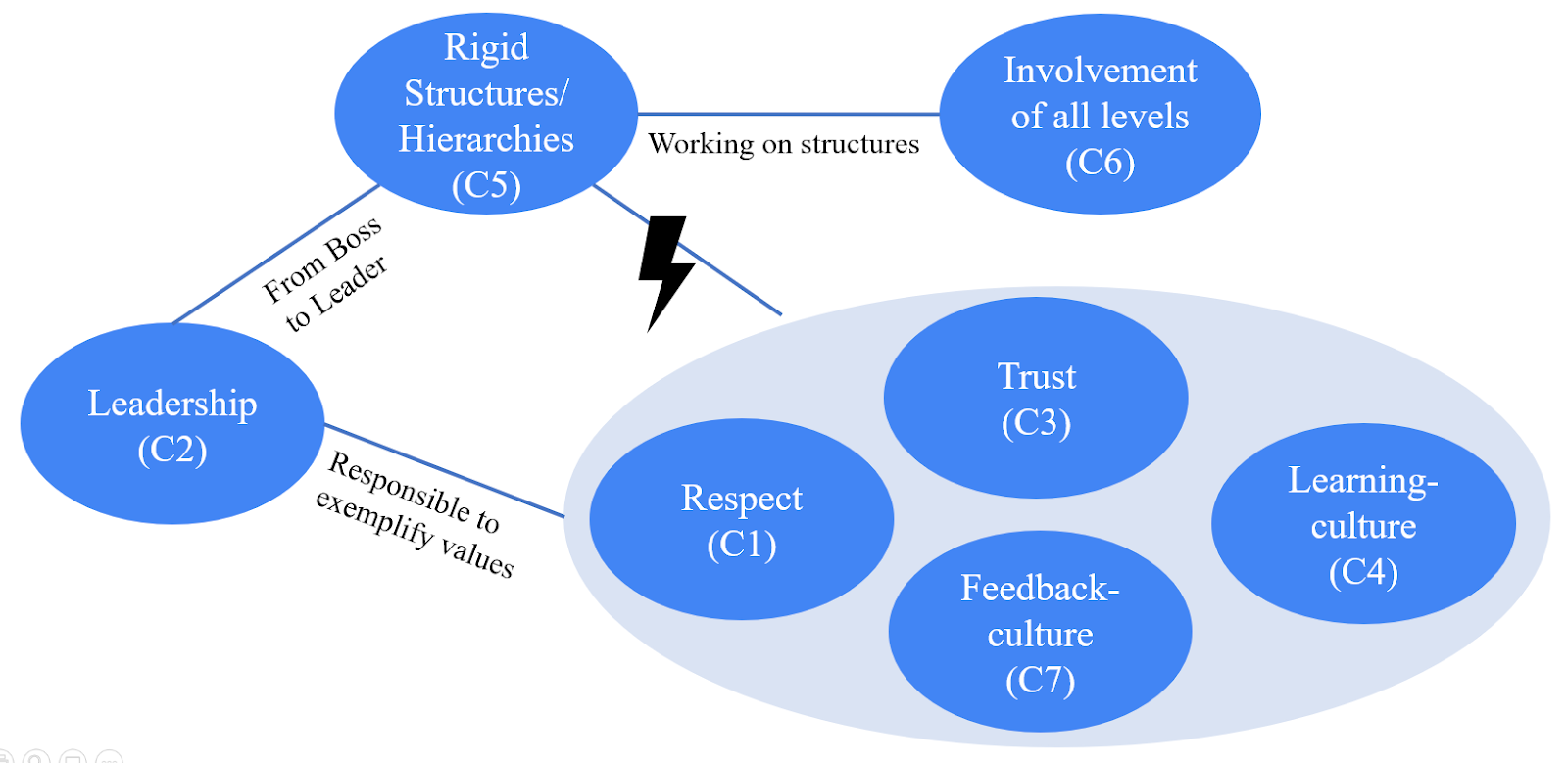}
\caption{Conceptual model of key challenges with agile culture}
\end{figure*}

\subsection{Implications on Practitioners and Researchers}
When we compare our results with the literature, it can be seen that cultural clashes and general organizational resistance to change are part of the most significant agile adoption barriers~\cite{VersionOne.2021}. Our key challenges with agile culture (see Table~\ref{tab3:Key Challenges with AC}) reveal that the main issues are in the area of respect (C1) and trust (C3, C4, and C7). These findings could be related to the fact that changes associated with (agile) transformation may not be sufficiently accompanied and could therefore fail~\cite{Kotter.1995}.

In this regard, our key challenges provide approaches to support practitioners and researchers with transformation processes. The key challenges can be used by practitioners as a specific tool to drive cultural change in an organization. For example, the key challenges help to raise awareness of problems that arise in the interaction between agility and organizational culture. The key challenges make it possible for humans within the organization not to rush into all problems at once, but to already have an initial approach as to which problems are most important. For instance, C6 (see Table~\ref{tab3:Key Challenges with AC}) shows that cultural change within an organization does not always progress at the same pace at the different levels (individual, team, organization). Practitioners should therefore consider how the key challenges manifest themselves at the different levels. Based on the analysis, solutions to the challenges can then be developed. Changing an organizational culture is complex and the change process should go through a series of phases that usually require a considerable length of time~\cite{Kotter.1995}. Therefore, appropriate models should be used if a change is to be initiated. The Cynefin framework~\cite{Snowden.2007} or the Deming circle~\cite{Deming.1986}, for example, are suitable for this purpose. In a complex context, appropriate steps can only be determined retrospectively. This requires an experimental approach (e.g. probe, sense, respond \cite{Snowden.2007}). Researchers can support agile transformation and cultural change initiatives in organizations, for instance, by developing appropriate tools. For example, there are already some tools available to measure agility and to support the agile transformation~\cite{Looks.2021,Patel.2009,Sidky.2007}. Based on these tools, another tool could be developed that focuses specifically on agile cultural change in an organization. Our identified key challenges could be used as a basis for this new tool.

\subsection{Limitations}
Although we designed and conducted our study based on established guidelines, there are some limitations which need to be taken into account.
\paragraph{Construct validity:} In our iterative research approach, we gathered feedback from the target group and did pretests at each step. In addition, we used a 7-point Likert scale in order to avoid interpolations in our quantitative survey. So, for each item the participants had the option to choose “no comment”.  
\paragraph{Internal validity:} The design of a questionnaire is important for the process of data collection. For this reason, we conducted several pretests of our questionnaire with participants who met the criteria of the target group. All participants completed the survey fully, so we were able to achieve a 0
\paragraph{External validity:} Both the qualitative survey and the quantitative survey were conducted in German language. We decided to design the data collection artifacts in German, because we wanted to address the study in our personal professional networks. Using our own network leads to several positive facets of conducting the study. For instance, we were able to collect data using a quantitative survey with a dropout rate of 0\%. However, the results may not be easily generalizable to other groups. Since the participants of the survey often work in the IT industry, we can assume that many of them work in an international context and therefore this limitation is not so strong. However, the general transferability of the results should be evaluated by conducting the survey in other languages in the future.

\section{Conclusion and Future Work}
\label{Sec5:ConclusionAndFutureWork}
This paper presents the results of our study addressing the key challenges that arise from the interplay between agility and organizational culture. Furthermore, we present a conceptual model that explains the relationships between the identified key challenges. Based on our results, we discuss practical implications aiming to present recommendations in order to provide important contributions to both practitioners and researchers in the area of agile software development and related fields that use agile way of working beyond IT.

We identified seven key challenges which emerge from the interplay of organizational culture and agile methods and practices. These key challenges are: respectful treatment between humans (C1), agile leadership (C2), trust in interactions (C3), learning culture (C4), rigid hierarchies (C5), involvement of all organizational levels (C6) and not valued feedback (C7). As mentioned above, we developed a conceptual model which represents the relations of the seven key challenges of an agile culture. The model shows that a rigid hierarchical structure hinders the development of an agile culture aiming to optimize aspects like value-based work or the learning culture. Also, the involvement of all organizational levels (management, team, individual) is important in order to optimize the structure as well as the culture of the organization. Finally, we encourage the management level of organizations to optimize themselves in order to be able to exemplify the relevant values for a successful usage of agile methods in practice. This transformation from a boss to a leader will support well-known success factors for agile software development teams \cite{Dikert.2016}, in particular the self-optimization which belongs to the feedback and learning culture of such teams.

Based on the results of our study, we provide practical implications aiming to give specific recommendations for a cultural change of organizations. Supporting the findings from other researchers (e.g., \cite{Diebold.2015,Gupta.2019,Tolfo.2011}), we strongly recommend the usage of existing transformation frameworks like the Cynefin framework or the Deming circle. 

However, the results of our study and other surveys (e.g., \cite{VersionOne.2021}) show the need for further research in the field of cultural influences on agile methods and practices. Due to the high degree of tailoring agile methods, we see a rise of complexity choosing the right approach in specific situations. Also, the different cultural aspects and influences described on various levels like the department or specific teams may lead to complex challenges in organizations. Thus, we call other researchers to study the relationship and interplay of cultural influences on the use and transition of agile methods in the area of software development and beyond.

\appendix
\section{Appendices}
\subsection{Challenges with Agile Culture}
The full data set including the 15 challenges with agile and organizational culture is shown in Table \ref{tab4:Appendix-Key Challenges with AC}.
\begin{table*}
 \caption{Appendix 1: Key Challenges with Agile Culture}
  \label{tab4:Appendix-Key Challenges with AC}
  \begin{tabular}{cp{0.4\linewidth}cp{0.05\linewidth}p{0.05\linewidth}p{0.05\linewidth}p{0.05\linewidth}cc}

\hline
ID & Item (EN) & Mean & Stand. deviation & Confid. (p=0.05) & Confid. intervall min. & Confid. intervall max. & totally important & N \\
\hline
C1 & Humans in an organization do not treat each other with respect. & 6.70 & 0.81 & 0.17 & 6.53 & 6.86 & 80.4\% & 92\\
\hline
C2 & Management expects change from employees without embodying agile values themselves. & 6.43 & 1.17 & 0.24 & 6.20 & 6.67 & 69.6\% & 92\\
\hline
C3 & The organizational culture does not create a context for trusting interactions. & 6.60 & 0.63 & 0.13 & 6.47 & 6.73 & 67.4\% & 92\\
\hline
C4 & Humans in an organization are not allowed to make mistakes. & 6.49 & 0.75 & 0.15 & 6.34 & 6.64 & 60.9\% & 92\\
\hline
C5 & The existing structures of an organization prevent an agile way of working. & 6.40 & 0.96 & 0.20 & 6.21 & 6.60 & 59.8\% & 92\\
\hline
C6 & Agile cultural change does not occur at all levels (individual, team, management) of the organization. & 6.23 & 1.12 & 0.23 & 6.00 & 6.47 & 53.3\% & 90\\
\hline
C7 & Feedback is not valued in an organization. & 6.35 & 0.76 & 0.16 & 6.19 & 6.50 & 51.1\% & 92\\
\hline
C8 & Humans in an organization do not want to change the way they work. & 6.16 & 1.03 & 0.21 & 5.94 & 6.37 & 46.7\% & 90\\
\hline
C9 & In an organization, there is no transparency regarding decisions. & 6.35 & 0.76 & 0.16 & 6.13 & 6.44 & 45.7\% & 92\\
\hline
C10 & In an organization, there is no transparency regarding processes. & 6.25 & 0.76 & 0.16 & 6.09 & 6.41 & 45.1\% & 91\\
\hline
C11 & Humans in an organization do not understand that agile transformation needs perseverance. & 6.13 & 0.94 & 0.19 & 5.94 & 6.33 & 44.4\% & 90\\
\hline
C12 & Humans in an organization don't want to leave their comfort zone. & 5.97 & 1.05 & 0.22 & 5.75 & 6.18 & 37.8\% & 90\\
\hline
C13 & There is no safe environment for experiments in the organization. & 5.97 & 0.87 & 0.18 & 5.79 & 6.15 & 30.8\% & 91\\
\hline
C14 & Humans in an organization have not internalized the difference between technical and cultural agility. & 5.73 & 1.18 & 0.24 & 5.49 & 5.97 & 30.3\% & 89\\
\hline
C15 & The cultural values of humans within an organization do not align with agile values. & 5.96 & 1.12 & 0.23 & 5.72 & 6.19 & 30.0\% & 90\\
\hline
\end{tabular}
\end{table*}

\begin{acks}
  The authors would like to thank all respondents of the study, who supported us in making these important results available for the practitioners and researchers community.

  The authors would also like to thank the Bagilstein company located in Main, Germany. The company supported the first author in providing an internship and the opportunity of gaining practical experiences in the field of agile software development. 

\end{acks}

\bibliographystyle{ACM-Reference-Format}
\bibliography{references} 

\end{document}